\documentclass[entropy,article,accept,moreauthors,pdftex,10pt,a4paper]{mdpi}
\pdfmapfile{+txfonts.map}
\usepackage{mathrsfs}
\usepackage{amsmath}
\usepackage[flushleft]{threeparttable}
\usepackage{amssymb}
\usepackage[figuresright]{rotating}
\usepackage{todonotes}
\usepackage{xcolor}
\usepackage{booktabs}
\usepackage{multirow}
\usepackage{soul}
\usepackage{microtype}

\usepackage{algorithmic}
\usepackage[lined,boxed,ruled,commentsnumbered]{algorithm2e}

\usepackage{upgreek}
\usepackage{textcomp}

\makeatletter      
                    \renewcommand{\maketag@@@}[1]{\hbox{\m@th\normalsize\normalfont#1}}%      
                    \makeatother
\firstpage{1}
\articlenumber{106}
\doinum{10.3390/e20020106}
\pubvolume{20}
\pubyear{2018}
\copyrightyear{2018}
\history{Received: 27 December 2017; Accepted: 30 January 2018; Published: 3 February 2018}

\Title{Point Divergence Gain and Multidimensional \\Data Sequences Analysis}
\Author{{Renata Rycht\'{a}rikov\'{a} $^{1,}$*,}
~Jan Korbel $^{2,3,4}$, Petr Mach\'{a}\v{c}ek $^1$ and Dalibor \v{S}tys $^1$}
\AuthorNames{{Renata Rycht\'{a}rikov\'{a},}~Jan Korbel, Petr Mach\'{a}\v{c}ek and Dalibor \v{S}tys}

\address{
$^1$\quad {Institute of Complex Systems}, South Bohemian Research Center of Aquaculture and Biodiversity of~Hydrocenoses, Kompetenzzentrum MechanoBiologie in Regenerativer Medizin, Faculty of Fisheries and Protection of Waters, University of South Bohemia in \v{C}esk\'{e} Bud\v{e}jovice, Z\'{a}mek 136, 373 33 Nov\'{e} Hrady, Czech Republic; info@imagecode.eu (P.M.); stys@frov.jcu.cz (D.\v{S}.)\\
$^2$ \quad Section for Science of Complex Systems, CeMSIIS, Medical University of Vienna, Spitalgasse 23, 1090~Vienna,~Austria; korbeja2@fjfi.cvut.cz\\
$^3$ \quad Complexity Science Hub Vienna, Josefst\"{a}dter Strasse 39, 1080 Vienna, Austria\\
$^4$ \quad Faculty of Nuclear Sciences and Physical Engineering, Czech Technical University in Prague, B\v{r}ehov\'{a} 7, 115~19~Prague, Czech Republic\\
}

\corres{Correspondence: rrychtarikova@frov.jcu.cz; Tel.: +420-387-773-833}

\abstract{We introduce novel information-entropic variables---a Point Divergence Gain ($\Omega_\alpha^{(l \rightarrow m)}$), a Point Divergence Gain Entropy ($I_{\alpha}$), and a Point Divergence Gain Entropy Density ($P_{\alpha}$)---which are derived from the R\'{e}nyi entropy and describe spatio-temporal changes between two consecutive discrete multidimensional distributions. The behavior of $\Omega_\alpha^{(l \rightarrow m)}$ is simulated for typical distributions and, together with $I_{\alpha}$ and $P_{\alpha}$, applied  in analysis and characterization of series of multidimensional datasets of computer-based and real images.}

\keyword{point divergence gain (PDG); R\'{e}nyi entropy; data processing}

\begin{document}
\section{Introduction}
\label{}
Extracting the information from raw data obtained from, e.g., a set of experiments, is a challenging~task. Quantifying the information gained by a single point of a time series, a pixel in an image, or a single measurement is important in understanding which points bring the most information about the underlying system. This task is especially delicate in case of time-series and image processing because the information is not only stored in the elements, but also in the interactions between successive points in a time series. Similar, when extracting information from an image, not~all pixels have the same information content. This type of information is sometimes called {local information} because the information depends not only on the frequency of the phenomenon but also on the position of the element in the structure. The most important task is to identify the sources of information and to quantify them. Naturally, it is possible to use standard data-processing techniques based on quantities from information theory like, e.g., Kullback--Leibler divergence.~On the other hand, the~mathematical rigorousness is typically compensated by an increased computational complexity. For~this end, a simple quantity called {Point Information Gain} and its relative macroscopic variables---a Point Information Gain Entropy and a Point Information Gain Entropy Density---were introduced in \cite{Ryc16b}. In \cite{Ryc16a}, mathematical properties of the Point Information Gain were extensively discussed and applications to real-image data processing were pointed out. From the mathematical point of~view, the Point Information Gain represents a change of information after removing an element of a particular phenomena from a distribution. The method is based on the R\'{e}nyi entropy, which has been already extensively used in multifractal analysis and data processing (see e.g., Refs. \cite{Ji14,Ryc16a,Ko14,Ji12} and references~therein).

In this article, we introduce an analogous variable to the Point Information Gain. This new variable locally determines an information change after an exchange of a given element in a discrete set. We use a simple concept of entropy difference between the original set and the set with the exchanged element. The resulting value is called {Point Divergence Gain} $\Omega_\alpha^{(l \rightarrow m)}$~\cite{Ryc15,Ryc16c}. The main idea is to describe the importance of changes in the series of images (typically representing a video record from an experiment) and extract the most important information from it. Similar to the Point Information Gain Entropy and the Point Information Gain Entropy Density, the macroscopic variables called a Point Divergence Gain Entropy $I_{\alpha}$ and a Point Divergence Gain Entropy Density $P_{\alpha}$ are defined to characterize subsequent changes in a multidimensional discrete distribution by one number. The~goal of this article is to examine and demonstrate some properties of these variables and use them for examination of time-spatial changes of information in sets of discrete multidimensional data, namely~series of images in image processing and analysis, after the exchange of a pixel of a particular intensity for a pixel at the same position in the consecutive image. The main reason for choosing the {Point Divergence Gain} as the relevant quantity for the analysis of spatio-temporal changes is the fact that it represents an information gain of each pixel change. One can also consider model-based approaches based on the theory of random-fields, which can be more predictive in some cases. On the other hand, the model-free approach based on entropy gives us typically more relevant information for real data,
where it is typically difficult to find an appropriate model.
For the overview of model-based approaches in the random field theory, one can consult, e.g., Refs. \cite{chev,eid,helle}.

The paper is organized as follows: in Section \ref{sec.2}, we define the main quantity of the paper, i.e., the~Point Divergence Gain and the related quantities and discuss its theoretical properties. In Section~\ref{sec.3}, we show applications of the Point Divergence Gain to image processing for both computer-based and real sequences of images. We show that the Point Divergence Gain can be used as a measure of difference for clustering methods and detects the most prominent behaviour of a system. In~Section~\ref{sec.4}, we explain the presented methods and finer technical details necessary for the analysis including~algorithms. Section~\ref{sec.5} is dedicated to conclusions. {All image data, scripts for histogram processing}, and Image Info Extractor Professional software for image processing are available via \url{sftp://160.217.215.193:13332/pdg} (user: anonymous; password: anonymous.).

\section{Basic Properties of Point Divergence Gain and Derived Quantities}\label{sec.2}
\unskip
\subsection{Point Divergence Gain}
\label{subsec:PDG}

Recently, a quantity called Point Information Gain (PIG, $\Gamma_{\alpha}^{(i)}$) \cite{Ryc15,Ryc16c} and its generalization based on the R\'{e}nyi entropy \cite{Ryc16a} have been introduced. We show how to apply the concept of PIG to sequence of multidimensional data frames.

Let us assume a set of variables with $k$ possible outcomes (e.g., possible colours of each pixel). The~$\Gamma_{\alpha}^{(i)}$ is a simple variable based on entropy difference and enables us to quantify an information gain of each phenomenon. It is simply defined as a difference between entropy of an original discrete~distribution
\begin{equation}\label{eq: or}
P = \{p_j\}_{j=1}^k = \left\{\frac{n_1}{n},\dots,\frac{n_k}{n}\right\},
\end{equation}
which typically describes a frequency histogram of possible outcomes. Let us also define a distribution, where one occurrence of the $i$-th phenomenon is omitted, i.e.,
\begin{equation}
P^{(i)} = \left\{p_j^{(i)}\right\}_{j=1}^k = \left\{\frac{n_1}{n-1},\dots,\frac{n_i-1}{n-1},\dots,\frac{n_k}{n-1}\right\}.
\end{equation}

Thus, the Point Information Gain is defined as
\begin{equation}
\Gamma^{(i)}_\alpha \equiv \Gamma^{(i)}_\alpha(P) = \mathscr{H}_\alpha\left(P^{(i)}\right) - \mathscr{H}_\alpha(P),
\end{equation}
where $\mathscr{H}_\alpha$ is the R\'{e}nyi entropy ({{Despite all computer
implementations} being calculated as $\log_2$, the~following derivations are written in
natural logarithm, i.e., $\ln$.})

\begin{equation}
\label{eq:Ren}
\mathscr{H}_\alpha(P) = \frac{1}{\alpha-1} \ln \sum_i p_i^{\alpha}.
\end{equation}

The R\'{e}nyi entropy represents a one-parametric class of information quantities tightly related to multifractal dynamics and enables us to focus on certain parts of the distribution \cite{Ji04a}.~Unlike~the typically used R\'{e}nyi's relative entropy~\cite{Re61,Ku57,Ji04a,Cs75,Ha06,Er07,Er10,Ji12,Ji14}, the Point Information Gain $\Gamma_{\alpha}^{(i)}$ is a simple, computationally tractable quantity. Its mathematical properties have been extensively discussed in \cite{Ryc16a}. On the same basis, we can define a Point Divergence Gain (PDG, $\Omega_\alpha^{(l \rightarrow m)}$), where a discrete distribution $P^{(i)}$ is replaced by a distribution
\begin{equation}\label{eq: rep}
P^{(l \rightarrow m)} = \left\{p_j^{(l \rightarrow m)}\right\}_{j=1}^k = \left\{\frac{n_1}{n},\dots,\frac{n_l-1}{n},\dots, \frac{n_m+1}{n},\dots,\frac{n_k}{n}\right\},
\end{equation}
which can be obtained from the original distribution $P$, where the occurrence of the examined $l$-th phenomenon ($n_l \in \mathbb{N}^+$) is removed and supplied by a point of the occurrence of the $m$-th phenomenon ($n_m \in \mathbb{N}_0$). The main idea behind the definition is to quantify the information change in the subsequent image, if only one point is changed. Analogous to the Point Information Gain $\Gamma_{\alpha}^{(i)}$, the Point Divergence Gain can be defined as

\begin{equation}
\Omega_\alpha^{(l \rightarrow m)} \equiv \Omega_\alpha^{(l \rightarrow m)}(P) = \mathscr{H}_\alpha\left(P^{(l \rightarrow m)}\right) - \mathscr{H}_\alpha(P).
\end{equation}

Let us first show its connection to the Point Information Gain $\Gamma_{\alpha}^{(i)}$. Since $P^{(l)} = P^{(l \rightarrow m,m)}$, it is possible to express the Point Divergence Gain as
\begin{equation}
\Omega_\alpha^{(l \rightarrow m)}(P)  = \mathscr{H}_\alpha\left(P^{(l \rightarrow m)}\right) - \mathscr{H}_\alpha\left(P^{(l \rightarrow m,m)}\right) + \mathscr{H}_\alpha\left(P^{(l)}\right)  - \mathscr{H}_\alpha(P) = \Gamma_\alpha^{(l)}(P) - \Gamma_\alpha^{(m)}(P^{(l \rightarrow m)}).
\end{equation}

Let us investigate mathematical properties of the PDG. The $\Omega_\alpha^{(l \rightarrow m)}$ can be rewritten as
\begin{eqnarray}
\label{eq:1}
\Omega_\alpha^{(l \rightarrow m)} &=& \mathscr{H}_\alpha\left(P^{(l \rightarrow m)}\right) - \mathscr{H}_\alpha(P)\nonumber\\
  &=& \frac{1}{1-\alpha}\ln \left(\sum_{j=1}^k \left(p_j^{(l \rightarrow m)}\right)^\alpha\right) - \frac{1}{1-\alpha} \ln \left(\sum_{j=1}^k {p_j^\alpha}\right) = \frac{1}{1-\alpha} \ln \left(\frac {\sum_{j=1}^k \left(p_j^{(l \rightarrow m)}\right)^\alpha}{\sum_{i=1}^k{p_j^\alpha}}\right).
\end{eqnarray}

By plugging the relative frequencies from Equations~\eqref{eq: or} and \eqref{eq: rep} into Equation~\eqref{eq:1}, we obtain
\begin{eqnarray}
\label{eq:4}
\Omega_\alpha^{(l \rightarrow m)} &=& \frac{1}{1-\alpha} \ln{\left[\frac{(n_{l}-1)^\alpha + (n_{m}+1)^\alpha + \sum_{j=1,j \neq l,m}^{k} {n_{j}^\alpha}}{\sum_{j=1}^{k}{n_{j}^\alpha}}  \right]} \nonumber \\
&=& \frac{1}{1-\alpha} \ln{\left[\frac{(n_{l}-1)^\alpha + (n_{m}+1)^\alpha + \sum_{j=1}^{k} {n_{j}^\alpha} - n_{l}^\alpha - n_{m}^\alpha}{\sum_{j=1}^{k}{n_{j}^\alpha}}  \right]} \nonumber \\ &=&
\frac{1}{1-\alpha} \ln{\left[\frac{(n_{l}-1)^\alpha - n_{l}^\alpha + (n_{m}+1)^\alpha - n_{m}^\alpha}{\sum_{j=1}^{k}{n_{j}^\alpha}} + 1 \right].}
\end{eqnarray}

As seen in Equation~\eqref{eq:4}, the variable $\Omega_\alpha^{(l \rightarrow m)}$ does not depend (contrary to the $\Gamma_\alpha^{(i)}$) on $n$ but depends only on the number of elements of each phenomenon $j$. In Equation~\eqref{eq:4}, let us design the nominator $\sum_{j=1}^k{n_{j}^\alpha}$, which is constant and related to the original distribution (histogram) of elements and to the parameter $\alpha$, as $\mathcal{C}_\alpha$. It gives us the final form\vspace{12pt}
\begin{equation}
\label{eq:6}
\Omega_\alpha^{(l \rightarrow m)} = \frac{1}{1-\alpha} \ln{\left[\frac{(n_{l}-1)^\alpha - n_{l}^\alpha + (n_{m}+1)^\alpha - n_{m}^\alpha}{\mathcal{C}_\alpha} + 1 \right].}
\end{equation}

Equation~\eqref{eq:6} demonstrates that, for a particular distribution, $\Omega_\alpha^{(l \rightarrow m)}$ is a function only of the parameter $\alpha$ and frequencies of occurrences of the phenomena $n_l$ and $n_m$ in the original distribution, between which the exchange of the element occurs.
%When $n_m, n_l \gg 1$, we can approximate $(x \pm 1)^\alpha \approx x^\alpha (1 \pm \alpha/x)$ and we obtain
%\begin{equation}\label{eq:7a}
%\Omega_\alpha^{(l \rightarrow m)}   \approx \frac{1}{1-\alpha} \ln\left(1 + \alpha \frac{\frac{1}{n_l}-\frac{1}{n_m}}{\mathcal{C}_\alpha} \right).
%\end{equation}
Equation~\eqref{eq:6} further shows that if the exchange of the element occurs between phenomena $l$ and $m$ of the same (similar) frequencies of occurrence (i.e.,~$n_l \approx n_m$), the value of $\Omega_\alpha^{(l \rightarrow m)}$ equals 0. If we remove a rare point and supply it by a high-frequency point (i.e., $n_l \ll n_m$), the value of $\Omega_\alpha^{(l \rightarrow m)}$ is negative, and \textit{vice versa}. Low values of parameter $\alpha$ separate low-frequency events as $\Omega_\alpha^{(l \rightarrow m)} = 0$, whereas high $\alpha$ emphasize high-frequency events as $\Omega_\alpha^{(l \rightarrow m)} \gg 0$ or $\Omega_\alpha^{(l \rightarrow m)} \ll 0$ and merge rare events into $\Omega_\alpha^{(l \rightarrow m)} = 0$. With respect to the previous discussion and practical utilization of this notion, we emphasize that, for real systems with large $n$, the $\Omega_\alpha^{(l \rightarrow m)}$ are rather small numbers.

\begin{figure}[H]
\centering
\includegraphics[width=15 cm]{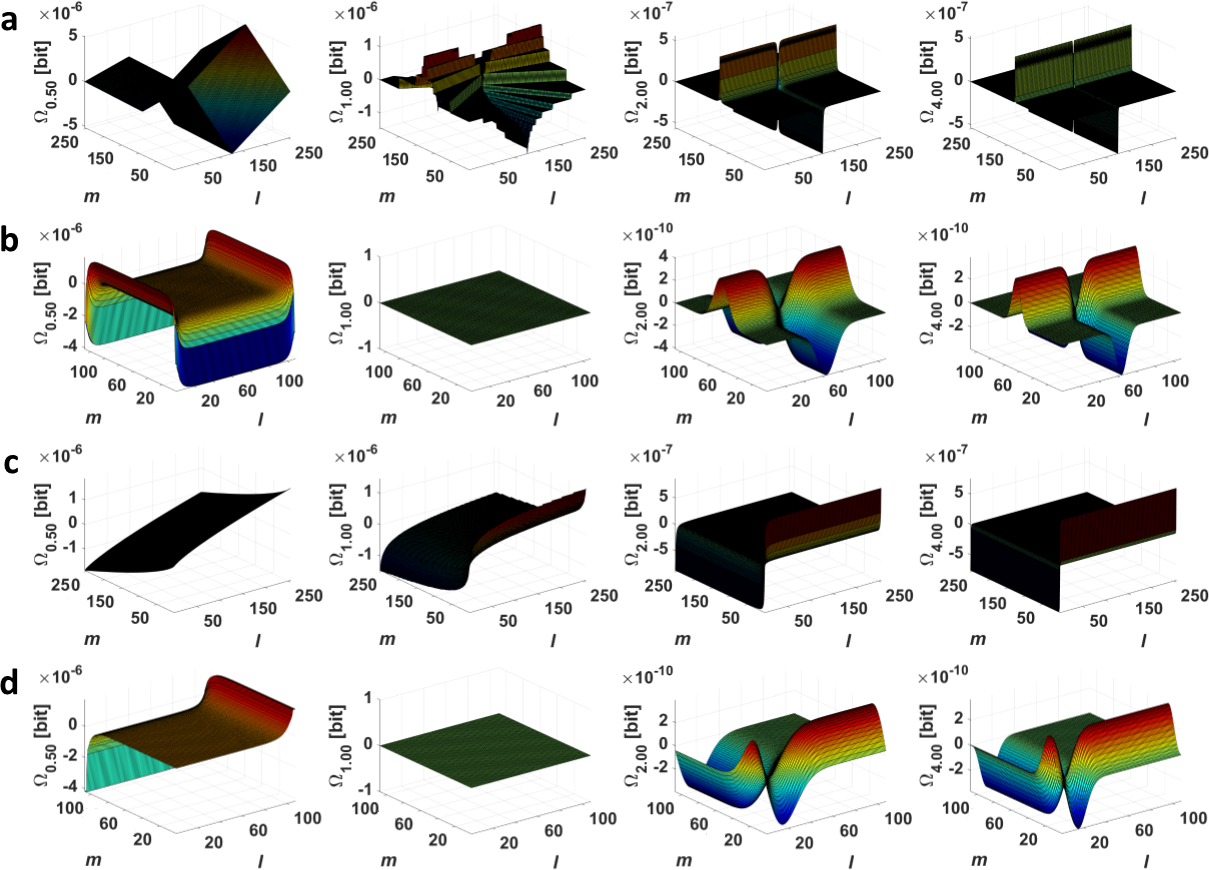}
\caption[]{The $\Omega_\alpha$-{transformations of the discrete} (\textbf{a}) Cauchy; (\textbf{b}) Gauss; (\textbf{c}) L\'{e}vy; and (\textbf{d}) Rayleigh distribution for $\alpha$ = $\{0.5; 1.0; 2.0; 4.0\}$ (Section~\ref{hist_process}).}
\label{Fig1}
\end{figure}

In the 3D plots of Figure~\ref{Fig1}, we demonstrate $\Omega_\alpha^{(l \rightarrow m)}$-transformations of four thoroughly studied distributions---the Cauchy, Gauss (symmetrical), L\'{e}vy, and Rayleigh distribution (asymmetric; all~specified in Section~\ref{hist_process})---for $\alpha = \{0.5; 1.0; 2.0; 4.0\}$, where each point presents the exchange of the element between bins $l$ and $m$ (Algorithm \ref{Alg1}). In this case, the (a)symmetry of the distribution is always maintained.

%The figures 1-7 you provided are still not very clear to be published, please improve the figure quality for these figures. You can print your figures as pdf file and enlarge to 400% to check if the figures are clear enough. In order not to infuluence the quality of your paper, please revise.
%The order figures 2 and 3 need to be revised.
%The figures 6 and 7 are reused and need to ask for copyright.

%%%%%%%%%%%%%%%%%%%%%%%%%%%%%%%%%%%%%%%%%%%%%%
\newpage
\begin{algorithm}[H]
\IncMargin{1em}
\LinesNumbered
\caption{Calculation of a point divergence gain matrix ($\mathbf{\Omega}_\alpha$) for typical histograms.\label{Alg1}}
%
%\begin{algorithmic}

\KwIn{$n$-bin histogram $\mathbf{h}$; $\alpha$, where $\alpha$ $\geq$ 0 $\land$ $\alpha$ $\neq$ 1}
\KwOut{$\mathbf{\Omega}_\alpha$}
\BlankLine
\BlankLine
$\mathbf{\Omega}_\alpha = $zeros($n,n$); \qquad \emph{\% create a zero square matrix $\mathbf{\Omega}_\alpha$ of the size of $n \times n$}\\
$\mathcal{C}_\alpha = \mbox{sum}(\mathbf{h}.^\land\alpha))$; \qquad \emph{\% calculate the constant $C_\alpha$ for the given distribution and $\alpha$}\

\BlankLine
\BlankLine
\For{$l = 1$ \KwTo $n$}{
\BlankLine
\BlankLine
\eIf{$\mathbf{h}(l) \neq 0$}{
\BlankLine
\BlankLine
\For{$l = 1$ \KwTo $n$}{
	$\mathbf{\Omega}_\alpha(l,m) = \log_2(((\mathbf{h}(l)-1)^\land\alpha - \mathbf{h}(l)^\land\alpha + (\mathbf{h}(m)+1)^\land\alpha - \mathbf{h}(m)^\land\alpha)/  \mathcal{C}_\alpha + 1)/(1-\alpha)$;
}
\BlankLine
\BlankLine
}{$\mathbf{\Omega}_\alpha$(l,:) = NaN;}
\emph{\% if the bin $l$ of the histogram $\mathbf{h}$ is occupied, calculate $\mathbf{\Omega}_\alpha$ at each position (l,m) according to Equation~\eqref{eq:6}, else set the not-a-number into the row l of the $\mathbf{\Omega}_\alpha$ matrix}\\
\BlankLine
\BlankLine
}
\BlankLine
\BlankLine
%\end{algorithmic}
\end{algorithm}
\BlankLine
\BlankLine

Now we will consider the specific case $\alpha=2$ (collision entropy) for which Equation~\eqref{eq:6} can be simplified to
\begin{equation}
\label{eq:8}
\Omega_2^{(l \rightarrow m)} = -\ln{\left[\frac{2}{\mathcal{C}_2}(n_m - n_l + 1) + 1\right]} = -\ln{\left[\frac{2}{\mathcal{C}_2}(\Delta n^{(l \rightarrow m)} + 1) + 1\right].}
\end{equation}	

For a specific difference $\Delta n^{(x \rightarrow y)} = D$, Equation~\eqref{eq:8} can be approximated by the 1st-order Taylor~sequence
\begin{eqnarray}
\label{eq:9}
\Omega_2^{(l \rightarrow m)} &\approx& -\ln{\left[\frac{2}{\mathcal{C}_2}(D + 1) + 1\right]} - \frac{2}{2(D + 1) + \mathcal{C}_2}(\Delta n^{(l \rightarrow m)} - D) \nonumber \\
&=& - \frac{2}{2D + 2 + \mathcal{C}_2}\Delta n^{(l \rightarrow m)} + \frac{2D}{2D + 2 + \mathcal{C}_2} -\ln{\left[\frac{2D}{\mathcal{C}_2} + {\mathcal{C}_2} + 1\right].}
\end{eqnarray}

\begin{figure}[H]
\centering
\includegraphics[width=\textwidth]{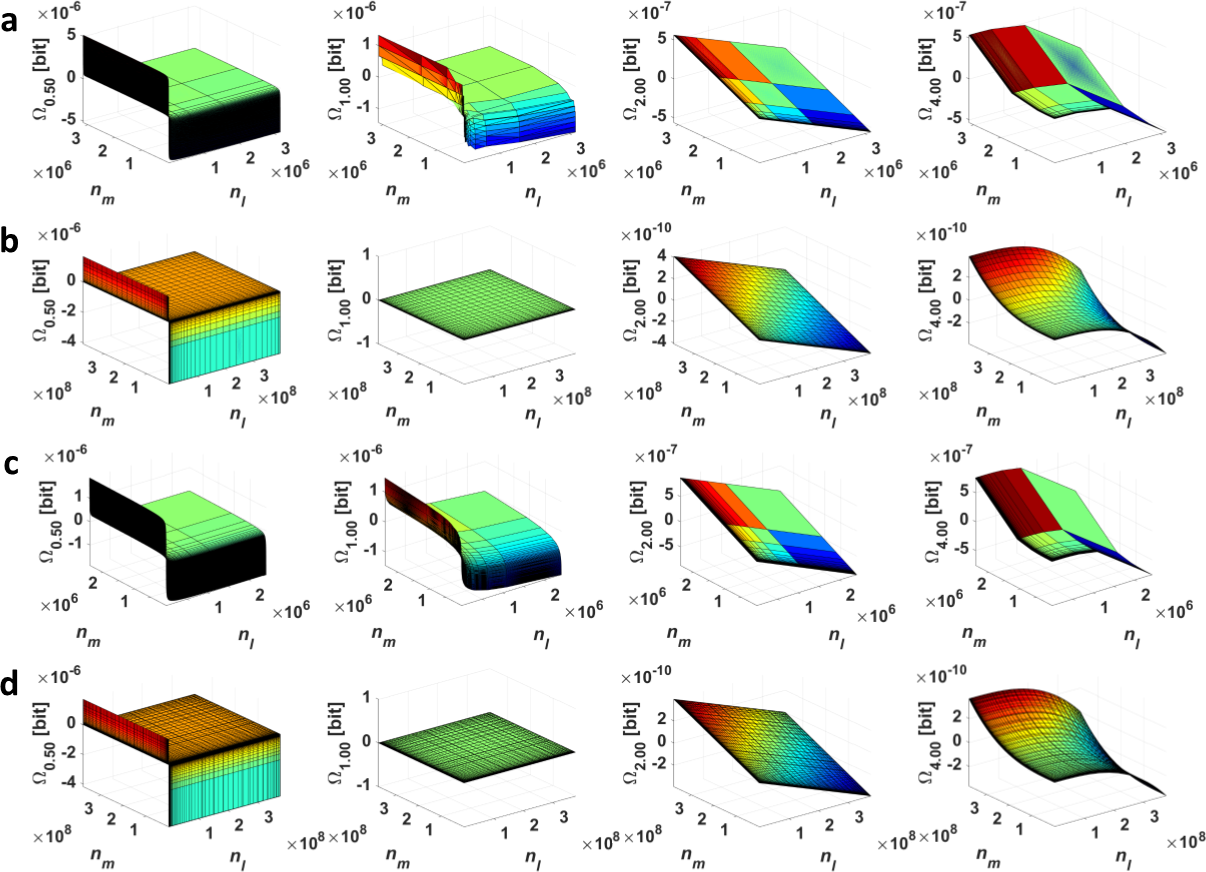}
\caption[]{{The dependencies} $\Omega_\alpha = f(n_m, n_m-n_l)$ for the discrete (\textbf{a}) Cauchy; (\textbf{b}) Gauss; (\textbf{c}) L\'{e}vy; and~(\textbf{d})~Rayleigh distribution at $\alpha$ = $\{0.5; 1.0; 2.0; 4.0\}$ (Section~\ref{hist_process}).}
\label{Fig2}
\end{figure}

Equations \eqref{eq:8} and \eqref{eq:9} show that, for each unique $\Delta n^{(x \rightarrow y)}$, the $\Omega_2^{(l \rightarrow m)}$ depends only on the difference between the bins $l$ and $m$, which the exchange of the element occurs between, and this dependence is almost linear. In other words, this explains why, for all distributions in {Figure}~\ref{Fig2}, the~dependencies $\Omega_2^{(l \rightarrow m)} = f(n_m, n_m-n_l)$ are planes.
%The figures should be refered in order. The figure 2 should be mentioned before figure 3. Please confirm and revise.

For $\alpha \rightarrow 1$, the R\'{e}nyi entropy becomes the ordinary Shannon entropy~\cite{Sh48} and we obtain (cf.~Equation~\eqref{eq:Ren})
\begin{eqnarray}
\mathscr{H}_1(P) = -\sum_{j=1}^{k}p_j\ln p_j = -\sum_{j=1}^{k}\frac{n_j}{n}\ln \frac{n_j}{n} = -\sum_{j=1,j\neq l,m}^{k}\frac{n_j}{n}\ln \frac{n_j}{n} - \frac{n_m}{n}\ln \frac{n_m}{n} - \frac{n_l}{n}\ln \frac{n_l}{n}
\end{eqnarray}
and
\begin{eqnarray}
\mathscr{H}_1(P^{(l \rightarrow m)}) = -\frac{n_m + 1}{n}\ln \frac{n_m + 1}{n} - \frac{n_l - 1}{n}\ln \frac{n_l - 1}{n} -\sum_{j=1,j\neq l,m}^{k}\frac{n_j}{n}\ln \frac{n_j}{n}.
\end{eqnarray}

The difference of these entropies (cf. Equation~\eqref{eq:4}) is gradually giving
\begin{small}\begin{eqnarray}
\label{eq:Sh}
\Omega_1^{(l \rightarrow m)} &=& -\frac{n_m + 1}{n}\ln \frac{n_m + 1}{n} - \frac{n_l - 1}{n}\ln \frac{n_l - 1}{n} + \frac{n_m}{n}\ln \frac{n_m}{n} + \frac{n_l}{n}\ln \frac{n_l}{n} \nonumber \\
&=& -\frac{n_m + 1}{n}\ln (n_m + 1) + \frac{n_m + 1}{n}\ln n - \frac{n_l - 1}{n}\ln (n_l - 1) + \frac{n_l - 1}{n}\ln n + \frac{n_m}{n}\ln n_m \nonumber \\
& & -\frac{n_m}{n}\ln n + \frac{n_l}{n}\ln n_l - \frac{n_l}{n}\ln n \nonumber \\
&=& \underbrace{(\frac{n_m + 1}{n} + \frac{n_l - 1}{n} - \frac{n_m}{n} - \frac{n_l}{n})}_{=0}\ln n - \frac{n_m}{n} \ln (n_m + 1) - \frac{1}{n} \ln (n_m + 1) - \frac{n_l}{n} \ln (n_l - 1) \nonumber \\
& & +\frac{1}{n} \ln (n_l - 1) + \frac{n_m}{n}\ln n_m + \frac{n_l}{n}\ln n_l \nonumber \\
&=& \frac{1}{n}(n_m \ln \frac{n_m}{n_m + 1} + n_l \ln \frac{n_l}{n_l - 1} + \ln \frac{n_l - 1}{n_m + 1}).
\end{eqnarray}\end{small}

\begin{figure}[H]
\centering
\includegraphics[width=15 cm]{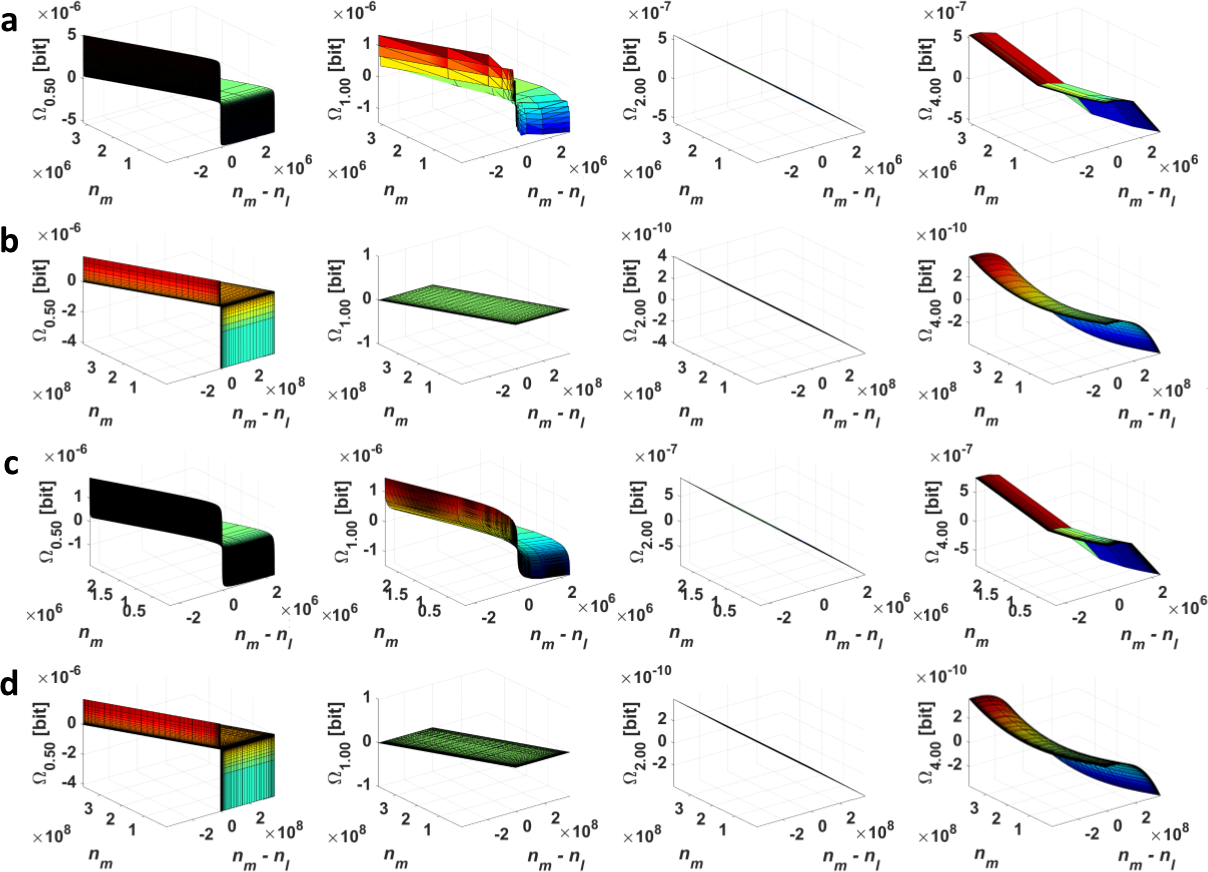}
\caption[]{{The dependencies} $\Omega_\alpha = f(n_l,n_m)$ for the discrete (\textbf{a}) Cauchy; (\textbf{b}) Gauss; (\textbf{c}) L\'{e}vy; and (\textbf{d})~Rayleigh distribution at $\alpha$ = $\{0.5; 1.0; 2.0; 4.0\}$ (Section~\ref{hist_process}).}
\label{Fig3}
\end{figure}

One can see that relation~(\ref{eq:Sh}) is defined for $n_l \in \mathbb{N}\setminus\{0,1\}$ and $n_m \in \mathbb{N}^+$ and is approximately equal to 0 for $n_l, n_m \gg 0$ (the Cauchy and Rayleigh distribution for $\alpha = 1$ in {Figure}~\ref{Fig3}).

For $n_l \in \mathbb{N}^+$ and $n_m \in \mathbb{N}_0$, from Equation~(\ref{eq:6}), further implies:
\begin{enumerate}[leftmargin=*,labelsep=4.9mm]
\item   If $\alpha = 0$, then $\Omega_0^{(l \rightarrow m)} = 0$.
\item   If $\alpha \rightarrow \infty$, then $\Omega_{\infty}^{(l \rightarrow m)} \rightarrow 0$.
\end{enumerate}

\subsection{Point Divergence Gain Entropy and Point Divergence Gain Entropy Density}
\label{subsec:entropies}
In this section, we introduce two new variables that help us to investigate changes between two (typically consecutive) points of time series. A typical example can be provided by video processing, where each element of a time or spatial series is represented by a frame. Let us have two data frames $\mathcal{I}_b = \{a_1,\dots,a_n\}$ and $\mathcal{I}_b = \{b_1,\dots,b_n\}$ ({{For simplicity, we use only one index which corresponds} to a one-dimensional frame. In case of images, we have typically two-dimensional frames and the elements are described by two indexes, e.g., $x$ and $y$ positions.}).
%Please confirm the footnotes.
At each position $i \in \{1,\dots,n\}$, it is possible to replace the value $a_i$ by the value of the following frame, i.e., $b_i$. The resulting $\Omega_\alpha^{(a_i \rightarrow b_i)}$ then quantifies how much information is gained/lost, when, at the $i$-th position, we replace the value $a_i$ for the value $b_i$. A Point Divergence Gain Entropy (PDGE, $I_\alpha$) is \textls[-5]{defined as a sum of absolute values of all PDGs for all~pixels,}~i.e.,

\begin{equation}
I_\alpha(\mathcal{I}_a;\mathcal{I}_b)= \sum_{i=1}^n |\Omega_\alpha^{(a_i \rightarrow b_i)}| =  \sum_{l=1}^{k} \sum_{m=1}^{k} n_{lm} |\Omega_\alpha^{(l \rightarrow m)}|,
\label{eq:13}
\end{equation}
where $n_{lm}$ denotes the number of present substitutions $l \rightarrow m$, when we transform $\mathcal{I}_a \rightarrow \mathcal{I}_b$. The~absolute value ensures that the contribution of the transformation of a rare point to a frequent point (negative $\Omega_\alpha$) and a frequent point to a rare point (positive $\Omega_\alpha$) do not cancel each other and both contribute to the resulting PDGE. Typically, appearance or disappearance of a rare point (and~replacement by a frequent value---typically background colour) carries important information about the experiment. The PDGE can be understood as an absolute information change.

Moreover, it is possible to introduce other macroscopic quantity---a Point Divergence Gain Entropy Density (PDGED, $P_\alpha$), where we do not sum over all pixels, but only over all realized transitions $l \rightarrow m$. Thus, the PDGED can be defined as
\begin{equation}
P_\alpha(\mathcal{I}_a;\mathcal{I}_b) = \sum_{l=1}^{k} \sum_{m=1}^{k} \chi_{lm} |\Omega_\alpha^{(l \rightarrow m)}|,
\label{eq:14}
\end{equation}
where
\begin{equation}
\chi_{lm} = \left\{
                          \begin{array}{ll}
                            1, & n_{lm} \geq 1,\\
                            0, & n_{lm} = 0.
                          \end{array}
                        \right.
\end{equation}

Let us emphasize that two transitions $a_1 \rightarrow b_1$ and $a_2 \rightarrow b_2$, where the frequencies of the occurrences of the phenomena $a_1$ and $a_2$ are equal and of the phenomena $b_1$ and $b_2$ are equal as well, give two unique values of the $\Omega_\alpha^{(a_i \rightarrow b_i)}$. In the computation of the PDGED, this is arranged by a hash function (Algorithm \ref{Alg2}). We can understand the quantity PDGED as an absolute information change of all realized transitions of phenomena $m \rightarrow l$.

\BlankLine
\BlankLine

%%%%%%%%%%%%%%%%%%%%%%%%%%%%%%%%%%%%%%%%%%%%%%
\begin{algorithm}[H]
\label{Alg2}
\IncMargin{1em}
\LinesNumbered
\caption{Calculation of a point information gain matrix ($\mathbf{\Omega}_\alpha$) and values $P_\alpha$ and $I_\alpha$ for two consecutive images of a time-spatial series.}
\KwIn{2 consecutive images $\mathbf{I_1}$ and $\mathbf{I_2}$ of the size $m \times n$; $\alpha$, where $\alpha$ $\geq$ 0 $\land$ $\alpha$ $\neq$ 1}
\KwOut{$\mathbf{\Omega}_\alpha$}
\BlankLine
\BlankLine
$\mathbf{h} = \mbox{hist}(\mathbf{I_1})$; \qquad \emph{\% create an intensity histogram $\mathbf{h}$ of the image $\mathbf{I_1}$}\\
$C_\alpha = \mbox{sum}(\mathbf{h}.^\land\alpha)$; \qquad \emph{\% calculate the constant $C_\alpha$ for the given distribution and $\alpha$}\\
$\mathbf{\Omega}_\alpha = \mathbf{I_1}.*0$; \qquad \emph{\% create a zero matrix $\mathbf{\Omega}_\alpha$ of the size of the $\mathbf{I_1}$}\

\textbf{hashMap} = containers.Map; \qquad \emph{\% declare an empty hash-map (the key-value array)}

\BlankLine
\BlankLine
\For{$i = 1$ \KwTo $(m \times n)$}{
$\mathbf{\Omega}_\alpha(i) = \log_2(((\mathbf{h}(\mathbf{I_1}(i+1))-1)^\land\alpha - \mathbf{h}(\mathbf{I_1}(i+1))^\land\alpha + (\mathbf{h}(\mathbf{I_2}(i+1)))^\land\alpha -$ \mbox{$ \mathbf{h}(\mathbf{I_2}(i+1))^\land\alpha)/\mathcal{C}_\alpha + 1)/(1-\alpha)$};\\
\qquad \emph{\% for each element $i$ of the image $\mathbf{I_1}$, calculate a value $\Omega_\alpha$ after replacement of the intensity \\ \qquad at the position $i$ in the histogram of image $\mathbf{I_1}$ by the intensity at the same position in the \\ \qquad image $\mathbf{I_2}$ (Equation~\eqref{eq:6})}

\BlankLine
\BlankLine

$v$ = \textbf{I}($i$);
\qquad \emph{\% read a value of the element (intensity) at the position i}\\

$checkSum$ = calcCheckSum(\textbf{h}, $v$); \\
\qquad \emph{\% calculate checkSum using a hash-function effective enough (e.g., MD4, MD5, SHA1)}\\

\BlankLine
\BlankLine

\If{\mbox{not} \textbf{\mbox{hashMap}}.isKey(checkSum)}
{\textbf{hashMap}($checkSum$) = $\mathbf{\Omega}_\alpha$(i);\\
\qquad \emph{\% if the hash-map does not contain the key, insert a new element with the key \\ \qquad checkSum, where the inserted value is the $\mathbf{\Omega}_\alpha$ at the position i}
}
}
\BlankLine
\BlankLine

$I_\alpha$ = sum(sum(abs($\mathbf{\Omega}_\alpha$)));\\
\qquad \emph{\% calculate I$_\alpha$ as a sum of all elements in the matrix $\mathbf{\Omega}_\alpha$ (Equation~\eqref{eq:13})}

$P_\alpha$ = sum(abs(values($\mathbf{hashMap}$)));\\
\qquad \emph{\% calculate P$_\alpha$ as a sum of all elements in the matrix $\mathbf{hashMap}$ (Equation~\eqref{eq:14})}

\BlankLine
\BlankLine
%\caption{Calculation of a point information gain matrix ($\mathbf{\Omega}_\alpha$) and values $P_\alpha$ and $I_\alpha$ for two consecutive images of a time-spatial series.}
\end{algorithm}
%%%%%%%%%%%%%%%%%%%%%%%%%%%%%%%%%%%%%%%%%%%%%%
\newpage
If the aim is to assess the influence of elements of a high occurrence on the time-spatial changes in the image series, it is recommended to use PDGE where each element is weighted by its number of occurrences. If the aim is to suppress the influence of these extreme values, it is better to compute~PDGED.

Let us consider a time-series $\mathcal{V}$, where each time step contains one frame, so $\mathcal{V} = \{\mathcal{I}_1,\mathcal{I}_2,\dots\}$. The series $\mathcal{V}$ can be, e.g., a sequence of images (a video) obtained from some experiment, etc. For~each time step, it is possible to calculate
$I_\alpha(t) = I_\alpha(\mathcal{I}_t;\mathcal{I}_{t+s})$, resp. $P_\alpha(t) = P_\alpha(\mathcal{I}_t;\mathcal{I}_{t+s})$, where $s$ is the time~lag. Typically, we assume $s=1$, i.e., consecutive frames with a constant time step.

\section{Application of Point Divergence Gain and Its Entropies in Image Processing}\label{sec.3}

The generalized Point Divergence Gain $\Omega_{\alpha}^{(l \rightarrow m)}$ in Equation~\eqref{eq:6} was originally used for characterization of dynamic changes in image series, namely in $z$-stacks of raw RGB data of unmodified live cells obtained via scanning along the $z$-axis using video-enhanced digital bright-field transmission microscopy~\cite{Ryc15,Ryc16c}. In these two references, this new mathematical approach utilizes 8- and 12-bit intensity histograms of two consecutive images for pixel-by-pixel intensity weighted (parameterized) subtraction of these images to suppress the camera-based noise and to enhance the image contrast ({{In~case of calibrated digital} camera-based images, where the value of each point of the image reflects a number of incident photons, or, in case of computer-based images, it can be sufficient to use a simple subtraction for evaluation of time-spatial changes in the image series.}).

For this paper, we chose other (grayscale) digital image series (Table \ref{Tab1}) in order to demonstrate other applications of the PDG mathematical approach in image processing and analysis. Moreover, we~newly introduce applications of the additive macroscopic variables Point Divergence Gain Entropy $I_\alpha$ and Point Divergence Gain Entropy Density $P_\alpha$.

\begin{table}[H]
\caption{Specifications of image series.\label{Tab1}}
\centering
%\begin{threeparttable}
\begin{tabular}{ll l l l l p{3cm}}
\toprule
\bfseries \bfseries Series & \bfseries Source & \bfseries Bit-Depth & \bfseries Number of Img. & \bfseries Resolution & \bfseries Origin\\
\midrule
Toy Vehicle & \cite{USC} & 8-bit & 10 & 512 $\times$ 512 & camera\\
Walter Cronkite & \cite{USC} & 8-bit & 16 & 256 $\times$ 256 & camera\\
Simulated BZ & \cite{Sty16b,Sty16c,Sty16d} & 8-bit & 10,521 & 1001 $\times$ 1001 & computer-based $^{a}$\\
Ring-fluorescence & & 12-bit & 1058 & 548 $\times$ 720 & experimental $^{b}$\\
Ring-diffraction & & 8-bit $^{c}$ & 1242 & 252 $\times$ 280 & experimental $^{b}$\\
\bottomrule
\end{tabular}

\begin{tabular}{@{}c@{}}
\multicolumn{1}{p{\textwidth -.88in}}{\footnotesize \textls[-15]{$^{a}$ A set of a noisy hotch-potch machine simulation of the Belousov--Zhabotinsky reaction~\cite{Sty16b,Sty16c,Sty16d} at 200~achievable states with the internal excitation of 10, and phase transition, internal excitation, and external neighbourhood kind of noise of 0, 0.25, and 0.15, respectively.} $^{b}$ The microscopic series of a 6-$\upmu$m standard microring (FocalCheck\textsuperscript{TM}, cat. No. F36909, Life Technologies\textsuperscript{TM} {(Eugene, OR, USA)}) were acquired using the CellObserver microscope (Zeiss, {Oberkochen,} Germany) at the EMBL (Heidelberg, Germany). For both light processes, the green region of the visible spectrum was selected using an emission and transmission optical filter, respectively. In case of the diffraction, the point spread function was separated and the background intensities was disposed using Algorithm \ref{Alg1} in~\cite{Ryc16c}. $^{c}$ The 12-bit depth was reduced using a Least Information Lost algorithm~\cite{Sty16}, which, by~shifting the intensity bins, filled all empty bins in the histogram obtained from the whole data series up and rescaled these intensities between their minimal and maximal value.}
\end{tabular}
\end{table}
%Please add.

\subsection{Image Origin and Specification}
\label{subsec:origin}
\begin{figure}[H]
\centering
\includegraphics[width=13 cm]{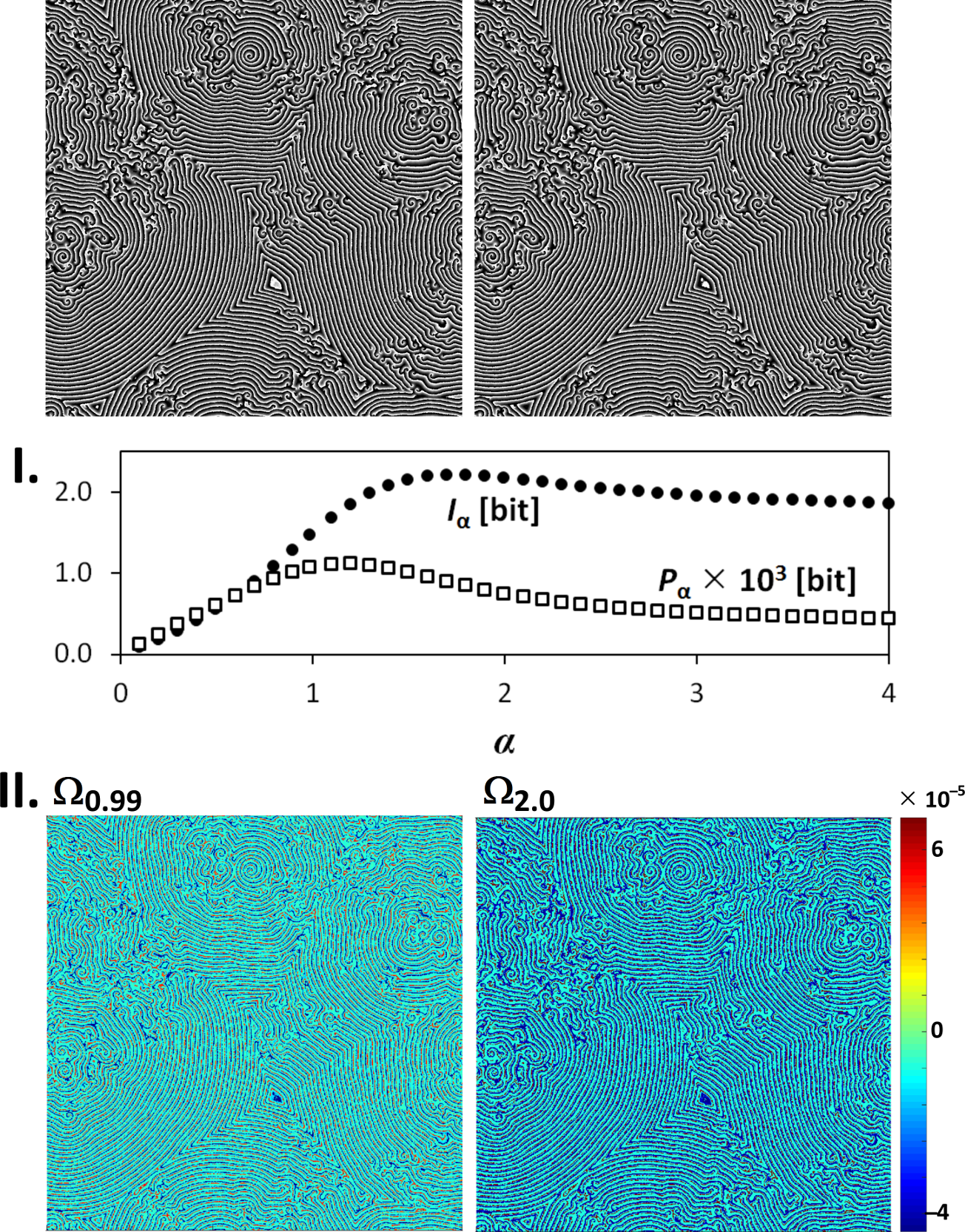}
\caption[]{{The} $I_\alpha$, $P_\alpha$, and $\Omega_{\alpha}$ for a pair of multifractal grayscale images. \textbf{I.} The $I_\alpha$ and $P_\alpha$ spectra, \textbf{II.}~8-bit visualization of $\Omega_{\alpha}$-values for $\alpha = \{0.99; 2.0\}$.}
\label{Fig5}
\end{figure}

\begin{figure}[H]
\centering
\includegraphics[width=11 cm]{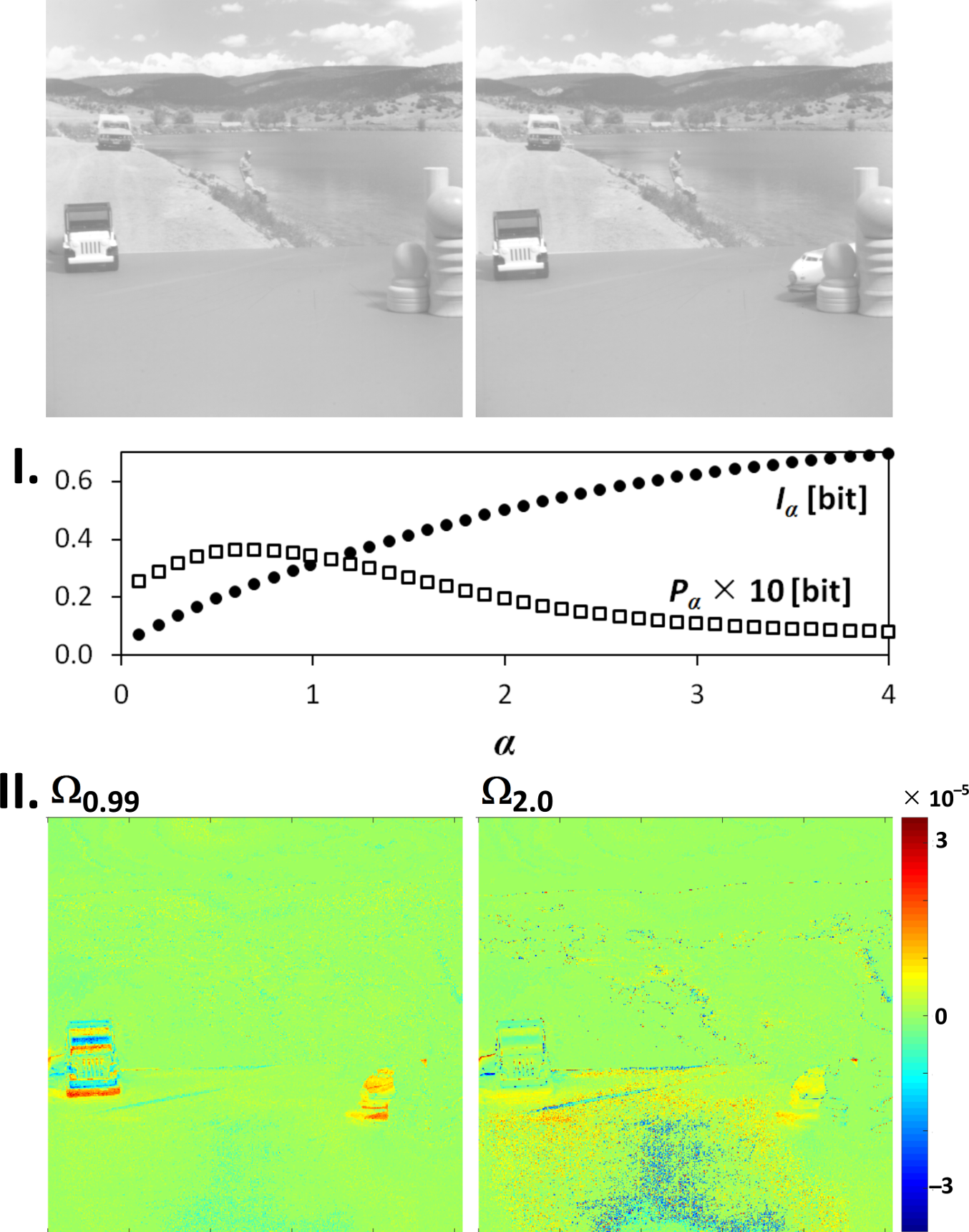}
\caption[]{{The} $I_\alpha$, $P_\alpha$, and $\Omega_{\alpha}$ for a pair of real-life grayscale images. \textbf{I.} the $I_\alpha$ and $P_\alpha$ spectra; \textbf{II.} 8-bit visualization of $\Omega_{\alpha}$-values for $\alpha = \{0.99; 2.0\}$.}
\label{Fig4}
\end{figure}

Owing to the relation of the $\Omega_\alpha^{(l \rightarrow m)}$ to the R\'{e}nyi entropy, the $I_\alpha$ and $P_\alpha$ as macroscopic variables can determine a fractal origin of images by plotting $I_\alpha = f_I(\alpha)$ and $P_\alpha = f_P(\alpha)$ spectra. If we deal with an image multifractality, the dependency $I_\alpha = f_I(\alpha)$ or the dependency $P_\alpha = f_P(\alpha)$ shows a~peak. In~case of a unifractality, these dependences are monotonous. It is demonstrated in Figures~\ref{Fig5} and \ref{Fig4}. There can be no doubts that the origin of the simulated Belousov--Zhabotinsky reaction (Figure~\ref{Fig5}) is multifractal. This statement is further strengthened by the courses of the dependencies $I_\alpha = f_I(\alpha)$ and $P_\alpha = f_P(\alpha)$, where we can see peaks with maxima at $\alpha \in (1,2)$. On the contrary, a pair of images in Figure~\ref{Fig4} (moving toys of cars) is a mixture of the objects of different fractal origin. In this case, whereas~the course of $f_I(\alpha)$ is monotonous and thus shows a unifractal characteristics, the dependence $f_P(\alpha)$ has a maximum at $\alpha = 0.6$ and thus demonstrates some multifractal features in the image. This~is due to the fact that, since each information contribution is counted only once, the $P_\alpha$ is more sensitive to the phenomena, which occur less frequently in the image. The monotonic course of the $P_\alpha$ would be achieved only when a sequence of time-evolved Euclidian objects was transformed into the~values~$\Omega_\alpha^{(l \rightarrow m)}$.

As mentioned in Section~\ref{subsec:entropies}, the variables $I_\alpha$ and $P_\alpha$ measure absolute information change between a pair of images and characterize a similarity between these images. Therefore, these variables can find a practical utilization in auto-focusing in both light and electron digital microscopy. The in-focus object can be defined as an image with the global extreme of $I_\alpha$ or $P_\alpha$. In other characteristics, this~image fulfils the Nijboer--Zernike definition~\cite{NZT}: it is the smallest and darkest image in light or electron diffraction or the smallest and brightest image in light fluorescence (Section~\ref{clustering}).

\subsection{Image Filtering and Segmentation}

Segmentation is a type of filtering of specific features in an image. The parameter $\alpha$ and the related value of $\Omega_\alpha^{(l \rightarrow m)}$ enable us to filter the parts of two consecutive images, which are either stable or differently variable in time. This can be employed in a 3D image reconstruction by thresholding and joining $\Omega_\alpha^{(l \rightarrow m)}$ = 0 from two consecutive images or in image tracking via thresholding of the highest and lowest $\Omega_\alpha^{(l \rightarrow m)}$ in a first image and the following image, respectively.

This is illustrated using simple examples in Figures~\ref{Fig5} and \ref{Fig4} where the highest (red-coded) and lowest (blue-coded) values of the $\Omega_\alpha^{(l \rightarrow m)}$ show the position of the object in the second and the first image of the image sequence, respectively. Compared with the $\Omega_{0.99}^{(l \rightarrow m)}$, the variance between the extremes of the $\Omega_{2.00}^{(l \rightarrow m)}$ is wider and the number of points $\Omega_{2.00}^{(l \rightarrow m)} = 0$ is lower.

In digital light transmission microscopy, this mathematical method enabled us to find time stable intracellular objects inside live mammalian cells from consecutive pixels that fulfilled the equality $\Omega_{\alpha}^{(l \rightarrow m)} = 0$ for $\alpha = 4.00$~\cite{Ryc15} or $\alpha = 5.00$~\cite{Ryc16c}. In these cases, the high value of $\alpha$ ensured merging rare points in the image, suppressing the camera noise that was reflected in the images and, thus, modelling the shape of organelles. The rest of image escaped the observation. In the next paper~\cite{RycBMC}, this method was extended to widefield fluorescent data.

As in the case of the Point Information Gain~\cite{Ryc16a}, the process of image segmentation of objects of a certain shape can be further improved by usage of the surroundings of this shape from which the intensity histogram is created for each pixel in the image.

\subsection{Clustering of Image Sets}
\label{clustering}
Finally, we used the Point Divergence Gain to detect the most relevant information contained in a sequence of images, capturing, e.g., an experiment. For this end, we used $I_\alpha$ or $P_\alpha$ as quantities of information change in the consecutive images and applied the clustering methods on them. The~values of $I_\alpha$ or $P_\alpha$ are small numbers (Section~\ref{subsec:PDG}). Due to the computation rounding of small numbers of the $I_\alpha$ and the $P_\alpha$ and for a better characterization of the image multifractality, in clustering, we use $\alpha$-dependent spectra of these variables than a sole number at one $\alpha$.

The dependence of the label of the cluster on the order of the image in the series is the smoothest for joint vectors $[I_\alpha, P_\alpha]$. The similarity of these vectors (and thus images as well) is described in a space of principal components, e.g.,~\cite{PCA}, and classified by standard clustering algorithms such as k-means++ algorithm~\cite{kmeans}. In comparison to the entropies and entropy densities related to the $\Gamma_\alpha^{(i)}$, the clustering using the $I_\alpha$ and the $P_\alpha$ is more sensitive to changes in the patterns (intensities) and does not require other specification of images by local entropies computed from a specific type of surroundings around each pixel.

\begin{figure}[H]
\centering
\includegraphics[width=12cm]{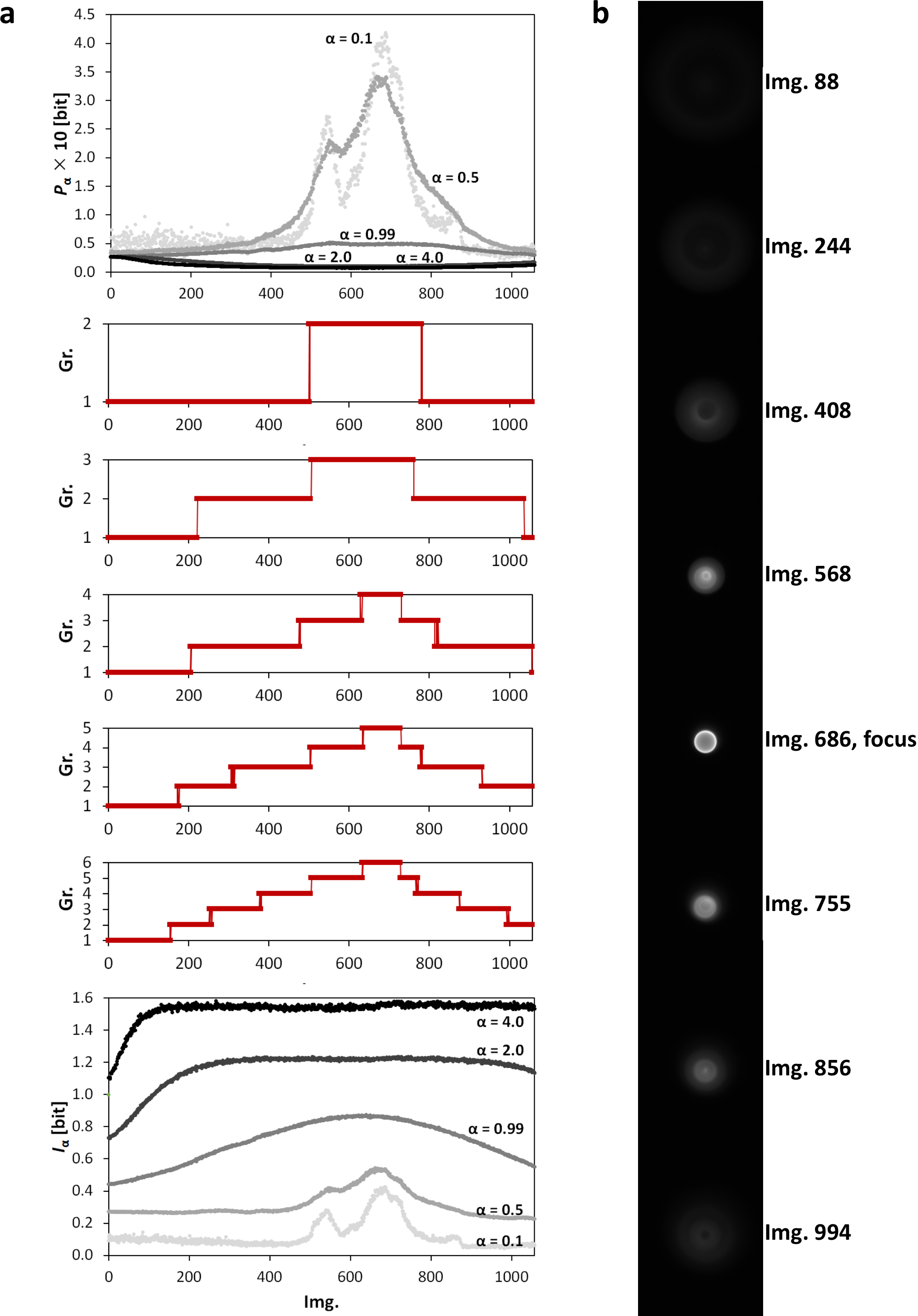}
\caption[]{{The results of clustering of a} $z$-stack of grayscale microscopic images of a microring obtained using a fluorescence mode. (\textbf{a}) the dependencies of ({upper}) the $P_\alpha$ and ({lower}) the $I_\alpha$ vs. order of the image in the $z$-stack for $\alpha = \{0.5; 0.99; 2.0; 4.0\}$ and ({middle}) clustering (k-means, squared Euclidian distance, 2--6 groups) of the $z$-stack using connected spectra [$I_\alpha$, $P_\alpha$] for \mbox{$\alpha = \{0.1; 0.3; 0.5; 0.7; 0.99; 1.3; 1.5; 1.7; 2.0; 2.5; 3.0; 3.5; 4.0\}$;} (\textbf{b}) the typical (middle) group's images for clustering into five groups (in (\textbf{a}), {middle}). The original 12-bit images are visualized in 8 bits using the Least Information Loss conversion~\cite{Sty16}.}
\label{Fig6}
\end{figure}

\begin{figure}[H]
\centering
\includegraphics[width=13 cm]{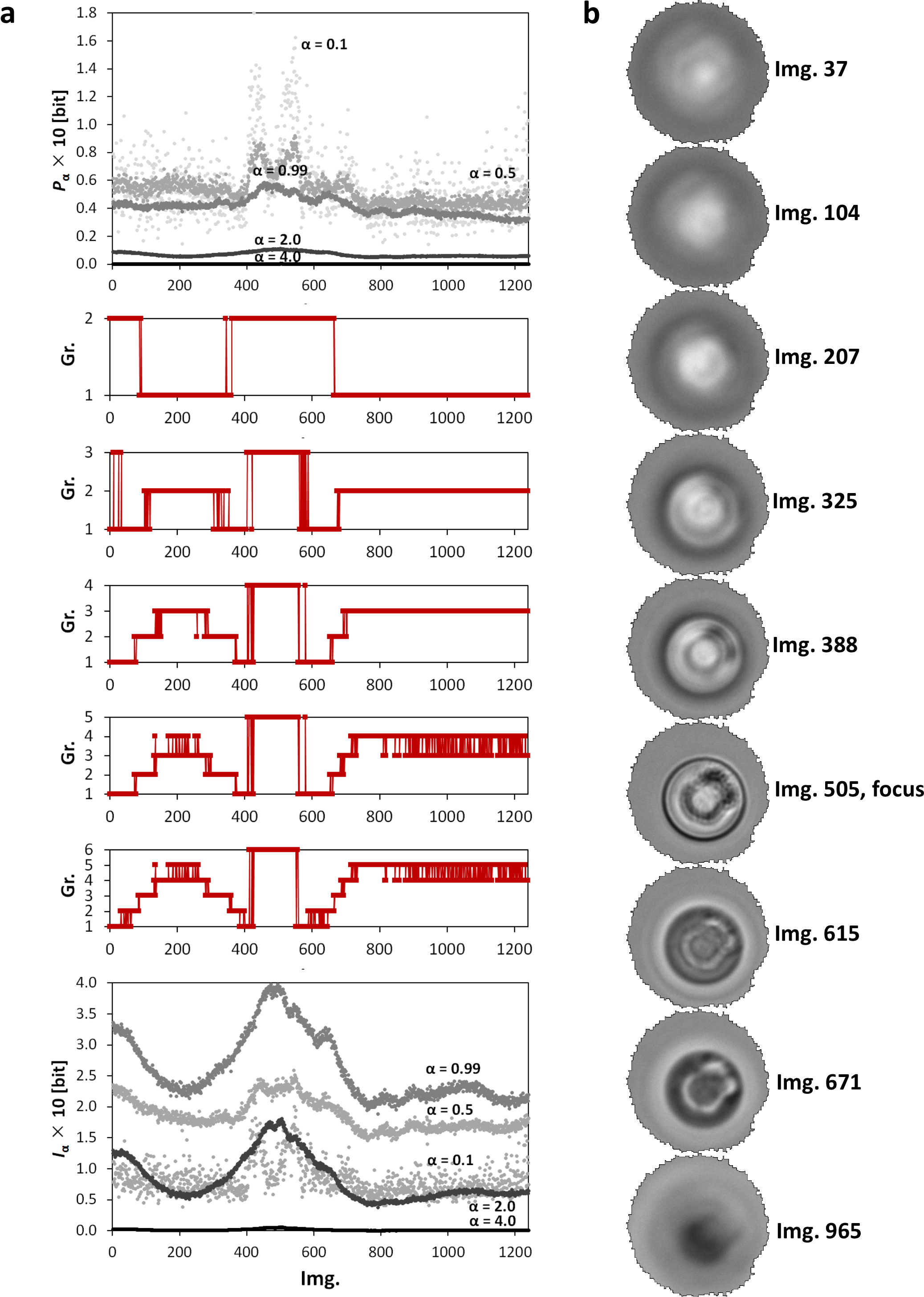}
\caption[]{{The results of clustering of a} $z$-stack of grayscale microscopic images of a microring obtained using a diffraction mode. (\textbf{a}) the dependencies of ({upper}) the $P_\alpha$ and ({lower}) the $I_\alpha$ vs. order of the image in the $z$-stack for $\alpha = \{0.5; 0.99; 2.0; 4.0\}$ and ({middle}) clustering (k-means, squared Euclidian distance, 2--6 groups) of the $z$-stack using connected spectra [$I_\alpha$, $P_\alpha$] for $\alpha = \{0.1; 0.3; 0.5; 0.7; 0.99; 1.3; 1.5; 1.7; 2.0; 2.5; 3.0; 3.5; 4.0\}$; (\textbf{b}) the typical (middle) group's images for clustering into 5 groups (in (\textbf{a}), {middle}). The original 12-bit images are visualized in 8 bits using the Least Information Loss conversion~\cite{Sty16}.}
\label{Fig7}
\end{figure}

The described clustering method was examined on $z$-stacks obtained using light microscopy. The~$z$-stacks were classified into 2--6 clusters (groups) when patterns of each image was described by 26~numbers, i.e., by vectors $[I_\alpha, P_\alpha]$ at 13 $\alpha$ (Figures~\ref{Fig6}a and \ref{Fig7}a). These clusters were evaluated on the basis of the sizes of intensity changes between images. These five classification graphs of the gradually splitting clusters (Figures~\ref{Fig6}a and \ref{Fig7}a, {middle}) further demonstrate the mutual similarity among the micrographs in each data series. The typical (middle) image of each cluster is shown in Figures~\ref{Fig6}b~and~\ref{Fig7}b.

Firstly, we shall deal with a $z$-stack with 1057 images of a microring obtained using a widefield fluorescent microscope. The results of clustering illustrate a canonically repetitive properties of the so-called point spread function as the image of the observed object goes to and from its focus. In~this~case, the image group containing the real focus of the maximal $I_\alpha$ and $P_\alpha$ at low $\alpha$ (Section~\ref{subsec:origin}) is successfully determined by clustering into two clusters (Figure~\ref{Fig6}a). However, we will aim for a description of the results for five clusters. The central Cluster 5 (94 images) can be called an object's focal region with image levels where parts of the object have their own focus. The in-focus cluster is asymmetrically surrounded by Cluster 4 (131 and 53 images below and above Cluster 5, respectively), which was set on the basis of the occurrence of the lower peaks of $I_\alpha$ and $P_\alpha$ at low $\alpha$. Cluster 3 (190~and 150 images below and above the focus, respectively) is typical of constant $I_\alpha$ and $P_\alpha$ for all $\alpha$. Cluster 2 contains img. 176--214 and the last 126 images. These images are characteristic of constant $I_\alpha$ and decreasing/increasing $P_\alpha$ at $\alpha \geq 2$. Cluster 1 (the first 175 images) is prevalently dominated by increasing $I_\alpha$ and decreasing $P_\alpha$ at high $\alpha$.

Before the calculation of the $I_\alpha$ and $P_\alpha$, the undesirable background intensities were removed from the images obtained using optical transmission microscopy. The rest of each image was rescaled into 8 bits (Section \ref{subsec:imageprocess}). The results of clustering of these images (Figure~\ref{Fig7}a) are similar to fluorescent data (Figure~\ref{Fig6}a). The light transmission point spread function is symmetrical around its focus as well but the pixels at the same $x,y$-positions below and above the focus have opposite, dark vs. bright, intensities. Furthermore, the transitional regions between the clusters are longer than for the fluorescent~data. The~central, in-focus, part of the $z$-stack (img. 427--561 in Cluster 4) with the highest peaks of $I_\alpha$ and $P_\alpha$ is unambiguously separated using four clusters. The focus itself lies at the 505th image. This~central part of the $z$-stack is surrounded by eight groups of images which were, due to their similarity, objectively classified into three clusters. Cluster 1 was formed by images 1--78, 376--426, and 562--661. These images show peaks of middle values of the $I_\alpha$ and $P_\alpha$. Images 79--153, 292--375, and 662--703 were classified into Cluster 2 (dominated by the local minimum of the $I_\alpha$ at $\alpha < 1$). Cluster~3 is related to the images with the lowest values of the $I_\alpha$ together with the lowest values and local peaks of the $P_\alpha$ for $\alpha < 1$ and for $\alpha < 1$, respectively. This cluster contains images 154--291 and the last 537~images of the series.

Let us mention that, in the clustering process, the $I_\alpha$ and $P_\alpha$ can recognize outliers such as  incorrectly saved images or images with illumination artifacts.

\section{Materials and Methods}\label{sec.4}
\label{methods}
\unskip
\subsection{Processing of Typical Histograms}
\label{hist_process}
For the Cauchy, L\'{e}vy, Gauss, and Rayleigh distributions, dependences of the $\Omega_{\alpha}^{(l \rightarrow m)}$ on the number of elements in bins $l$ and $m$ were calculated for $\alpha$ = $\{$0.1, 0.3, 0.5, 0.7, 0.99, 1.3, 1.5, 1.7, 2.0, 2.5, 3.0, 3.5, 4.0$\}$ using a pdg$\_$histograms.m {Matlab$^\circledR$ 2014 script} ({Mathworks, Natick, MA,} ~USA). The~following probability density functions $f(x)$ were studied:
%Please provide the version of Matlab
%it is a company, please provide the city and abbreviation of the state

\begin{enumerate}[leftmargin=*,labelsep=4.9mm]
\item   L\'{e}vy distribution:
\begin{equation}
f(x) = \mbox{round}\left[10^c\frac{\exp\left({-\frac{1}{2x}}\right)}{\sqrt{2\pi x^3}}\right], \quad x \in \mathbb{N}, \quad
\begin{cases}
x \in [1, 256],\quad c \in \{5,7\},\\
x \in [1,85],\quad c = 3,
\end{cases}
\end{equation}

\item   Cauchy distribution:
\begin{equation}
f(x) = \mbox{round}\left[10^c\frac{1}{\pi \left(1+x^2\right)}\right], \quad x \in \mathbb{Z}, \quad \begin{cases}
x \in [-127, 127],\quad c = 7,\\
x \in [-44, 44],\quad c = 3.5,
\end{cases}
\end{equation}

\item   Gauss distribution:
\begin{equation}
f(x) = \mbox{round}\left[10^c\frac{\exp\left({-\frac{x^2}{2\sigma^2}}\right)}{\sigma \sqrt{2\pi}}\right], \quad x \in \mathbb{Z}, \quad
\begin{cases}
x \in [-4, 4],\quad c = 4,\quad \sigma = 1, \\
x \in [-29, 29],\quad c = 3,\quad \sigma = 10, \\
x \in [-36, 36],\quad c = 4,\quad \sigma = 10, \\
x \in [-64, 64],\quad c = 10,\quad \sigma = 10,
\end{cases}
\end{equation}

\item Rayleigh distribution:
\begin{equation}
f(x) = \mbox{round}\left[10^c\frac{x}{b^2}\exp\left(-\frac{x^2}{2b^2}\right)\right], \quad x \in \mathbb{N}, \quad x \in [1, 108], \quad c = 10, \quad b = 16.
\end{equation}
\end{enumerate}

In Figure~\ref{Fig1}, the Cauchy and L\'{e}vy distributions at $c$ = 7 and the Gauss distribution at parameters $c$~=~10 and $\sigma$ = 10 are depicted.

\subsection{Image Processing and Analysis}
\label{subsec:imageprocess}
Image analysis based on calculation of the $\Omega_\alpha^{(l \rightarrow m)}$, $I_\alpha$, and $P_\alpha$ is demonstrated on five standard grayscale multi-image series (Table~\ref{Tab1}). All images were processed using {Whole Image} mode in an Image Info Extractor Professional software ({Institute of Complex Systems},%please provide the company and city
~ FFPW, USB, Nov\'{e} Hrady, Czech Republic). A~pair of images 5000--5001 of a simulated Belousov--Zhabotinsky (BZ) reaction and a pair of images {motion01.512}--{motion02.512}
%Should there be a space after "motion" here?
were recalculated for 40 values of $\alpha$ = $\{$0.1, 0.2, ..., 0.9, 0.99, 1.1, 1.2, ...,~4.0$\}$. The rest of series were processed for 13 values of $\alpha$ = $\{$0.1, 0.3, 0.5, 0.7, 0.99, 1.3, 1.5, 1.7, 2.0, 2.5, 3.0, 3.5,~4.0$\}$. The transformation at 13 $\alpha$ was followed by clustering of the matrices \mbox{[$P_\alpha$, $I_\alpha$]} vs. Img. by $k$-means method (squared Euclidian distance metrics). Due to a high data variance in the BZ simulation, the clustering was preceded by the z-score standardization of the matrices over $\alpha$. The~resulted indices of clusters were reclassified to be consecutive (i.e., the first image of the series and the first image of the following group are classified into gr. 1 and 2, respectively, etc.).

%\begin{tablenotes}
%			\item[a] \footnotesize{A set of a noisy hotch-potch machine simulation of the Belousov-Zhabotinsky reaction~\cite{Sty16b,Sty16c,Sty16d} at 200 achievable states with the internal excitation of 10, and phase transition, internal excitation, and external neighbourhood kind of noise of 0, 0.25, and 0.15, respectively.\\}
%			\item[b] \footnotesize{The microscopic series of a 6-$\mu$m standard microring (FocalCheck\textsuperscript{TM}, cat. No. F36909, Life Technologies\textsuperscript{TM}) were acquired using the CellObserver microscope (Zeiss, Germany) at the EMBL (Heidelberg, Germany). For both light processes, the green region of the visible spectrum was selected using an emission and transmission optical filter, respectively. In case of the diffraction, the point spread function was separated and the background intensities was disposed using Algorithm 1 in~\cite{Ryc16c}\\}
%			\item[c] \footnotesize{The 12-bit depth was reduced using a Least Information Lost algorithm~\cite{Sty16}, which, by shifting the intensity bins, filled all empty bins in the histogram obtained from the whole data series up and rescaled these intensities between their minimal and maximal value.\\}
%            \end{tablenotes}
%     \end{threeparttable}
%\end{center}

%\subsection{\highlight{Algorithms}}
%Algorithms must be close to where it was first mentioned, Algorithms 1 and 2 are fist cited in section 2, please revise and confirm if the section 4.3 is necessary
\label{sec:algorithms}
%The $\Omega_\alpha^{(l \rightarrow m)}$, $I_\alpha$, and $P_\alpha$ for all typical histograms and images were computed using Equations~\eqref{eq:6}, \eqref{eq:13}, and~\eqref{eq:14}. The implementation in the Image Info Extractor Professional is part of Algorithms~\ref{Alg1}~and~\ref{Alg2}. The double precision floating point data format of the $\Omega_{\alpha}$ is represented by rescaling into 8 bits. Let us note that, for $\alpha$ = 1, the software switches to the calculation of the Shannon entropy~\cite{Sh48}.\\

\section{Conclusions}\label{sec.5}

In this paper, we derived novel variables from the R\'{e}nyi entropy---a Point Divergence Gain $\Omega_\alpha^{(l \rightarrow m)}$, a Point Divergence Gain Entropy $I_\alpha$, and a Point Divergence Gain Entropy Density $P_\alpha$. We~have discussed their theoretical properties and made a brief comparison with the related quantity called Point Information Gain $\Gamma_\alpha^{i}$~\cite{Ryc16a}. Moreover, we have shown that the $\Omega_\alpha^{(l \rightarrow m)}$ and related quantities can find their applications in multidimensional data analysis, particularly in video processing. However, due to element-by-element computation, we can characterize time-spatial (4-D) changes much more sensitively than using, e.g., the previously derived $\Gamma_\alpha^{i}$. The $\Omega_\alpha^{(l \rightarrow m)}$ can be considered as a microstate of the information changes in the space-time. However, the $\Omega_\alpha^{(l \rightarrow m)}$, $I_\alpha$, and $P_\alpha$ show a property that is similar to the $\Gamma_\alpha^{i}$ and its relative macroscopic variables. Due to the derivation from the R\'{e}nyi entropy, they are good descriptors of multifractility. Therefore, they can be utilized to characterize patterns in datasets and to classify the (sub)data into groups of similar properties. This has been successfully utilized in clustering of multi-image sets, image filtration, and image segmentation, namely in microscopic digital imaging.

%\supplementary{All processed data, algorithms, and a Image Info Extractor Professional software are available via \url{sftp://160.217.215.193:13332/pdg} (user: anonymous; password: anonymous.)
\vspace{6pt}
\acknowledgments{This work was supported by the Ministry of Education, Youth and Sports of the Czech Republic---projects CENAKVA (No. CZ.1.05/2.1.00/01.0024), CENAKVA II (No. LO1205 under the {NPU} I program), the CENAKVA Centre
Development (No. CZ.1.05/2.1.00/19.0380)---and from the European Regional Development Fund in frame of the project Kompetenzzentrum MechanoBiologie (ATCZ133) in the Interreg V-A Austria---Czech Republic programme. J.K. acknowledges the financial support from the Czech  Science Foundation Grant No. 17-33812L and the Austrian Science Fund, Grant No. I 3073-N32.}

\authorcontributions{Renata Rycht\'{a}rikov\'{a} is the main author of the text and tested the algorithms; Jan Korbel is responsible for the theoretical part of the article; Petr Mach\'{a}\v{c}ek is the developer of the Image Info Extractor Professional software; Dalibor \v{S}tys is the group leader who derived the point divergence gain. All authors have read and approved the final manuscript.}

\conflictsofinterest{The authors declare no conflict of interest.}

\bibliographystyle{plain}
\reftitle{References}

\end{document}